\documentstyle[12pt,epsf]{article}

\renewcommand{\bar}[1]{\overline{#1}}
\newcommand{\longvec}[1]{\overrightarrow{\!\!#1}}
\newcommand{\etal}{{\em et al.}}
\newcommand{\ie}{{\it i.e.}}
\newcommand{\eg}{{\it e.g.}}
\begin{document}
\begin{flushright}
SLAC--PUB--7152\\
April 1996
\end{flushright}
\bigskip\bigskip

\thispagestyle{empty}
\flushbottom

\centerline{\Large\bf LIGHT-CONE QUANTIZATION AND} 
\smallskip
\centerline{{\Large\bf HADRON STRUCTURE}
    \footnote{\baselineskip=14pt
     Work supported by the Department of Energy, contract 
     DE--AC03--76SF00515.}}
\vspace{22pt}
  \centerline{\bf Stanley J. Brodsky}
\vspace{8pt}
  \centerline{\it Stanford Linear Accelerator Center}
  \centerline{\it Stanford University, Stanford, California 94309}
  \centerline{e-mail: sjbth@slac.stanford.edu}
\vspace*{0.9cm}

\vfill
\centerline{\large Talk presented at the }
\centerline{\large International Conference on Orbis Scientiae 1996}
\centerline{\large Miami Beach, Florida -- January 25--28, 1996}
\vfill
\newpage
\section{INTRODUCTION}

Quantum chromodynamics provides a fundamental
description of hadronic and nuclear structure and dynamics in terms
of elementary quark and gluon degrees of freedom. In practice,
the direct application of QCD to reactions involving the structure
of hadrons is extremely complex because of the interplay of
nonperturbative effects such as color confinement and multi-quark
coherence.  
 
In this talk, I will discuss light-cone quantization and the
light-cone  Fock expansion as a tractable and consistent
representation of relativistic many-body systems and bound states in
quantum field theory.  The Fock state representation in QCD 
includes all quantum fluctuations of the hadron wavefunction,
including far off-shell configurations such as intrinsic strangeness
and charm and,  in the case of nuclei, hidden color. The Fock state
components of the hadron with small transverse size, which dominate
hard exclusive reactions, have small color dipole moments and thus
diminished hadronic interactions. Thus QCD predicts minimal
absorptive corrections, \ie, color transparency for quasi-elastic
exclusive reactions in nuclear targets at large momentum transfer. 
In other applications, such as the calculation of the axial,
magnetic, and quadrupole moments of light nuclei, the QCD
relativistic Fock state description provides new insights which go
well beyond the usual assumptions of traditional hadronic and
nuclear physics.

\section{QCD ON THE LIGHT CONE }

The bound state structure of hadrons plays a critical role in
virtually every area of particle physics phenomenology.  For
example, in the case of the nucleon form factors, pion
electroproduction $ep \rightarrow e \pi^+n$, exclusive $B$ decays,
and open charm photoproduction $\gamma p\rightarrow D\Lambda_c$, the
cross sections depend not only on the nature of the quark currents,
but also on the coupling of the quarks to the initial and final
hadronic states.  Exclusive decay amplitudes such as $B \rightarrow
K^*\gamma$, processes which will be studied intensively at $B$
factories, depend not only on the underlying weak transitions
between the quark flavors, but also the wavefunctions which describe
how the $B$ and $K^*$ mesons are assembled in terms of their
fundamental quark and gluon constituents.  Unlike the leading twist
structure functions measured in deep inelastic scattering, such
exclusive channels are sensitive to the structure of the hadrons at
the amplitude level and to the coherence between the contributions
of the various quark currents and multi-parton amplitudes.

The analytic problem of describing QCD bound states is compounded
not only by the physics of confinement, but also by the fact that
the wavefunction of a composite of relativistic constituents has to
describe systems of an arbitrary number of quanta with arbitrary
momenta and helicities.  The conventional Fock state expansion based
on equal-time quantization quickly becomes intractable because of
the complexity of the vacuum in a relativistic quantum field theory.
Furthermore, boosting such a wavefunction from the hadron's rest
frame to a moving frame is as complex a problem as solving the bound
state problem itself.  The Bethe-Salpeter bound state formalism,
although manifestly covariant, requires an infinite number of
irreducible kernels to compute the matrix element of the
electromagnetic current even in the limit where one constituent is
heavy.

Light-cone quantization (LCQ) is formally similar to equal-time
quantization (ETQ) apart from the choice of initial-value surface. 
In ETQ one chooses a surface of constant time in some Lorentz frame
on which to specify initial values for the fields.  In quantum field
theory this corresponds to specifying commutation relations among
the fields at some fixed time. The equations of motion, or the
Heisenberg equations in the quantum theory, are then used to evolve
this initial data in time, filling out the solution at all spacetime
points.

In LCQ one chooses instead a hyperplane tangent to the light
cone---properly called a null plane or light front---as the
initial-value surface.  To be specific, we  introduce LC coordinates
\begin{equation}
x^\pm \equiv x^0\pm x^3
\end{equation}
(and analogously for all other four-vectors).  The selection of the
3 direction in this definition is of course arbitrary.
In terms of LC coordinates, a contraction of four-vectors decomposes
as
\begin{equation}
p\cdot x = \frac{1}{2}(p^+x^-+p^-x^+)-p_\perp \cdot x_\perp\; ,
\end{equation}
from which we see that the momentum ``conjugate'' to $x^+$ is $p^-$.
Thus the operator $P^-$ plays the role of the Hamiltonian in this
scheme, generating evolution in $x^+$ according to an equation of
the form (in the Heisenberg picture) 
\begin{equation} 
[\phi,P^-] = 2i\, \frac{\partial\phi}{\partial x^+}\; . 
\end{equation}

As was first shown by Dirac \cite{dirac49}, seven of the ten
Poincar\'e generators become kinematical on the LC , the maximum
number possible. The most important point is that these include
Lorentz boosts.  Thus in the LC representation boosting states is
trivial---the generators are diagonal in the Fock representation so
that computing the necessary exponential is simple.  One result of
this is that the LC theory can be formulated in a manifestly
frame-independent way, yielding wavefunctions that depend only on
momentum fractions and which are valid in any Lorentz frame.  This
advantage is somewhat compensated for, however, in that certain
rotations become nontrivial in LCQ.  Thus rotational invariance will
not be manifest in this approach.

Another advantage of going to the LC is even more striking: the
vacuum state seems to be much simpler in the LC representation than
in ETQ.  Note that the longitudinal momentum $p^+$ is conserved in
interactions.  For particles, however, this quantity is strictly
positive,
\begin{equation} 
p^+=\left(p_3^2+p_\perp^2+m^2\right)^{\frac{1}{2}} + p^3 > 0\; . 
\end{equation} 
Thus the Fock vacuum is the only state in the theory with $p^+=0$,
and so it must be an exact eigenstate of the full interacting
Hamiltonian. Stated more dramatically, the Fock vacuum in the LC
representation is the {\em physical} vacuum state. To the extent
that this is really true, it represents a tremendous simplification,
as attempts to compute the spectrum and wavefunctions of some
physical state are not complicated by the need to recreate a ground
state in which processes occur at unrelated \hbox{locations} and energy
scales.  Furthermore, it immediately gives a constituent picture;
all the quanta in a hadron's wavefunction are directly connected to
that hadron.  This allows a precise definition of the partonic
content of hadrons and makes interpretation of the LC wavefunctions
unambiguous.  It also raises the question, however, of whether LC
field theory can be equivalent in all respects to field theories
quantized at equal times, where nonperturbative effects often lead
to nontrivial vacuum structure.  In QCD, for example, there is an
infinity of possible vacua labelled by a continuous parameter
$\theta$, and chiral symmetry is spontaneously broken.  The question
is how it is possible to identify and incorporate such phenomena
into a formalism in which the vacuum state is apparently simple.

The description of relativistic composite systems using light-cone
quantization \cite{dirac49} thus appears to be  remarkably simple. 
The Heisenberg problem for QCD can be written in the form
\begin{equation}
H_{LC }\vert H\rangle = M_H^2 \vert H\rangle\; ,
\end{equation}
where $H_{LC}=P^+ P^- - P_\perp^2$ is the mass operator.  The
operator $P^-=P^0-P^3$ is the generator of translations in the
light-cone time $x^+=x^0+x^3.$ The quantities $P^+=P^0+P^3$ and
$P_\perp$ play the role of the conserved three-momentum.  Each
hadronic eigenstate $\vert H\rangle$ of the QCD light-cone
Hamiltonian can be expanded on the complete set of eigenstates
$\{\vert n\rangle\} $ of the free Hamiltonian which have the same
global quantum numbers: $\vert H\rangle=\sum\psi^H_n(x_i, k_{\perp
i}, \lambda_i) \vert n\rangle.$ In the case of the proton, the Fock
expansion begins with the color singlet state $\vert u u d \rangle $
of free quarks, and continues with $\vert u u d g \rangle $ and the
other quark and gluon states that span the degrees of freedom of the
proton in QCD.  The Fock states $\{\vert n\rangle \}$ are built on
the free vacuum by applying the free light-cone creation operators.
The summation is over all momenta $(x_i, k_{\perp i})$ and
helicities $\lambda_i$ satisfying momentum conservation $\sum^n_i
x_i = 1$ and $\sum^n_i k_{\perp i}=0$ and conservation of the
projection $J^3$ of angular momentum.

The wavefunction $\psi^p_n(x_i, k_{\perp i},\lambda_i)$ describes
the probability amplitude that a proton of momentum $P^+= P^0+P^3$
and transverse momentum $P_\perp$ consists of $n$ quarks and gluons
with helicities $\lambda_i$ and physical momenta $p^+_i= x_i P^+$
and $p_{\perp i} = x_i P_\perp + k_{\perp i}$.  The wavefunctions
$\{\psi^p_n(x_i, k_{\perp i},\lambda_i)\},n=3,\dots$ thus describe
the proton in an arbitrary moving frame.  The variables $(x_i,
k_{\perp i})$ are internal relative momentum coordinates.  The
fractions $x_i = p^+_i/P^+ = (p^0_i+p^3_i)/(P^0+P^3)$, $0 <x_i <1$,
are the boost-invariant light-cone momentum fractions; $y_i= \log
x_i$ is the difference between the rapidity of the constituent $i$
and the rapidity of the parent hadron.  The appearance of relative
coordinates is connected to the simplicity of performing Lorentz
boosts in the light-cone framework.  This is another major advantage
of the light-cone representation.

The  spectra of hadrons and nuclei as well as their scattering
states can be identified with the set of eigenvalues of the
light-cone Hamiltonian $H_{LC}$ for QCD.  Particle number is
generally not conserved in a relativistic quantum field theory, so
that each eigenstate is represented as a sum over Fock states of
arbitrary particle number. Thus in QCD each hadron is expanded as
second-quantized sums over fluctuations of color-singlet quark and
gluon states of different momenta and number. The coefficients of
these fluctuations are the light-cone wavefunctions $\psi_n(x_i,
k_{\perp i}, \lambda_i).$ The invariant mass ${\cal M}$ of the
partons in a given $n$-particle Fock state can be written in the
elegant form
\begin{equation} 
{\cal M}^2 = \sum_{i=1}^n \frac{k_{\perp i}^2+m^2}{x_i}\; . 
\end{equation} 
The dominant configurations in the wavefunction are generally those
with minimum values of ${\cal M}^2$.  Note that, except for the case
where $m_i=0$ and $k_{\perp i}=0$, the limit $x_i\rightarrow 0$ is
an ultraviolet limit, \ie, it corresponds to particles moving with
infinite momentum in the negative $z$ direction: $k^z_i\rightarrow -
k^0_i \rightarrow - \infty.$ The light-cone wavefunctions encode the
properties of the hadronic wavefunctions in terms of their quark and
gluon degrees of freedom, and thus all hadronic properties can be
derived from them.  The natural gauge for light-cone Hamiltonian
theories is the light-cone gauge $A^+=0$.  In this physical gauge
the gluons have only two physical transverse degrees of freedom, and
thus it is well matched to perturbative QCD calculations.

Since QCD is a relativistic quantum field theory, determining the
wavefunction of a hadron is an extraordinarily complex
nonperturbative relativistic many-body problem. In principle it is
possible to compute the light-cone wavefunctions by diagonalizing
the QCD light-cone Hamiltonian on the free Hamiltonian basis.  In
the case of QCD in one space and one time dimensions, the
application of discretized light-cone quantization (DLCQ)
\cite{bp91} provides complete solutions of the theory, including the
entire spectrum of mesons, baryons, and nuclei, and their
wavefunctions \cite{burkardt89,hbp90}.  In the DLCQ method, one
simply diagonalizes the light-cone Hamiltonian for QCD on a
discretized Fock state basis. The DLCQ solutions can be obtained for
arbitrary parameters including the number of flavors and colors and
quark masses.  More recently, DLCQ has been applied to new variants
of QCD$_{1+1}$ with quarks in the adjoint representation, thus
obtaining color-singlet eigenstates analogous to gluonium states
\cite{dkb94}.

The extension of this program to physical theories in 3+1 dimensions
is a formidable computational task because of the much larger number
of degrees of freedom; however, progress is being made.  Analyses of
the spectrum and light-cone wavefunctions of positronium in
QED$_{3+1}$ are given elsewhere \cite{kpw92}.  Hiller, Okamoto and I 
\cite{hbo94} have been  pursuing a nonperturbative calculation of
the lepton anomalous moment in QED using the DLCQ method. Burkardt
has recently solved scalar theories with transverse dimensions by
combining a Monte Carlo lattice method with DLCQ \cite{burkardt94}. 
Also of interest is recent work of Hollenberg and Witte
\cite{Hollenberg}, who have shown how Lanczos tri-diagonalization
can be combined with a plaquette expansion to obtain an analytic
extrapolation of a physical system to infinite volume.

There has also been considerable work on the truncations required to
reduce the space of states to a manageable level
\cite{phw90,perry94,wwhzgp94}.  The natural language for this
discussion is that of the renormalization group, with the goal being
to understand the kinds of effective interactions that occur when
states are removed, either by cutoffs of some kind or by an explicit
Tamm-Dancoff truncation.  Solutions of the resulting effective
Hamiltonians can then be obtained by various means, for example
using DLCQ or basis function techniques.  Some calculations of the
spectrum of heavy quarkonia in this approach have recently been
reported \cite{martina}.

One of the remarkable simplicities of the LC formalism is the fact
that one can write down exact expressions for the spacelike
electromagnetic form factors $\langle P+Q\vert J^+ \vert P\rangle$
of any hadrons for any initial or final state helicity.   At a fixed
light-cone time, the exact Heisenberg current can be identified with
the free current $j^+$.  It is convenient to choose the frame in
which $q^+=0$ so that $q_\perp^2$ is $Q^2 = -q_\mu^2.$ Since the
quark current $j^+$ has simple matrix elements between free Fock
states, each form factor for a given helicity transition
$\lambda\rightarrow \lambda^\prime$ can be evaluated from simple
overlap integrals of the light-cone wavefunctions \cite{dy70,bd80}:
\begin{equation} 
F_{\lambda^\prime, \lambda}(Q^2)= \sum_n\int \prod
d^2 k_{\perp i} \int{\prod d x_i} \overline
\psi_{n,\lambda^\prime}(x_i, k^{\prime}_{\perp i},\lambda_i)
\psi_{n,\lambda}(x_i,k_{\perp i},\lambda_i)\; , 
\end{equation} 
where the integrations are over the unconstrained relative
coordinates. The internal transverse momenta of the final state
wavefunction are $k^{\prime}_\perp = k_\perp + (1-x) q_\perp$ for
the struck quark and $k_\perp^{\prime} = k_\perp -x q_\perp$ for the
spectator quarks. Thus given the light-cone wavefunctions
$\{\psi_n(x_i, k_{\perp_i},\lambda_i)\}$ one can compute the
electromagnetic and weak form factors from a simple overlap of
light-cone wavefunctions, summed over all Fock states
\cite{dy70,bd80}. For spacelike momentum transfer only diagonal
matrix elements in particle number $n^\prime = n$ are needed. In
contrast, in the equal-time theory one must also consider
off-diagonal matrix elements and fluctuations due to particle
creation and annihilation in the vacuum.  In the nonrelativistic
limit one can make contact with the usual formulae for form factors
in Schr\"odinger many-body theory.

The structure functions of a hadron can be computed from the square
integral of its LC wavefunctions \cite{lb80}.  For example, the
quark distribution measured in deep inelastic scattering at a given
resolution $Q^2$ is
\begin{equation}
q(x_{Bj},Q^2)= \sum_n \int^{k_\perp^2 <Q^2} \prod d^2 k_{\perp i}
\int{\prod d x_i} \vert\psi_n(x_i, k_{\perp i},\lambda_i)\vert^2
\delta(x_q=x_{Bj})\; ,
\end{equation}
where the struck quark is evaluated with its light-cone fraction
equal to the Bjorken variable: $x_q = x_{Bj}=Q^2/2 p \cdot q.$ A
summation over all contributing Fock states is required to evaluate
the form factors and structure functions. Thus the hadron and
nuclear structure functions are the probability distributions
constructed from integrals over the absolute squares $\vert \psi_n
\vert^2 $, summed over $n.$ In the far off-shell domain of large
parton virtuality, one can use perturbative QCD to derive the
asymptotic fall-off of the Fock amplitudes, which then in turn leads
to the QCD evolution equations for distribution amplitudes and
structure functions. More generally, one can prove factorization
theorems for exclusive and inclusive reactions which separate the
hard and soft momentum transfer regimes, thus obtaining rigorous
predictions for the leading power behavior contributions to large
momentum transfer cross sections.  One can also compute the far
off-shell amplitudes within the light-cone wavefunctions where heavy
quark pairs appear in the Fock states.  Such states persist over a
time $\tau \simeq P^+/{\cal M}^2$ until they are materialized in the
hadron collisions.  As we shall discuss below, this leads to a
number of novel effects in the hadroproduction of heavy quark
hadronic states \cite{bhmt92}.

Although we are still far from solving QCD explicitly, a number of
properties of the light-cone wavefunctions of the hadrons are known
from both phenomenology and the basic properties of QCD.  For
example, the endpoint behavior of light-cone wavefunctions and
structure functions can be determined from perturbative arguments
and Regge arguments.  Applications are presented elsewhere
\cite{bbs94}. There are also correspondence principles. For example,
for heavy quarks in the nonrelativistic limit, the light-cone
formalism reduces to conventional many-body Schr\"odinger theory. 
On the other hand, one  can also build effective three-quark models
which encode the static properties of relativistic baryons.

\section{SOLVING NONPERTURBATIVE QUANTUM FIELD THEORY USING LCQ}

A large number of studies have been performed of model field
theories in the LC framework.  This approach has been remarkably
successful in a range of toy models in 1+1 dimensions: Yukawa theory
\cite{pb85}, the Schwinger model (for both massless and massive
fermions) \cite{schwinger,mccartor91}, $\phi^4$ theory \cite{hv8x},
QCD with various types of matter
\cite{burkardt89,hbp90,dkb94,klebanov,Anton}, and the sine-Gordon
model \cite{burkardt93}.  It has also been applied with promising
results to theories in 3+1 dimensions, in particular QED
\cite{kpw92} and Yukawa theory \cite{yuk}.  In all cases agreement
was found between the LC calculations and results obtained by more
conventional approaches, for example, lattice gauge theory.  In many
cases the physics of spontaneous symmetry breaking and vacuum
structure of the equal-time theory is represented by the physics of
zero modes in LCQ \cite{BR95}.

\subsection{ QCD$_{1+1}$ with Fundamental Matter }

This theory was originally considered by 't Hooft in the limit of
large $N_c$ \cite{thooft74}.  Later Burkardt \cite{burkardt89}, and
Hornbostel, Pauli and I \cite{hbp90}, gave essentially complete
numerical solutions of the theory for finite $N_c$, obtaining the
spectra of baryons, mesons, and nucleons and their wavefunctions. 
The results are consistent with the few other calculations available
for comparison, and are generally much more efficiently obtained. 
In particular, the mass of the lowest meson agrees to within
numerical accuracy with lattice Hamiltonian results \cite{hamer}. 
For $N_c=4$ this mass is close to that obtained by 't Hooft in the
$N_c\rightarrow\infty$ limit \cite{thooft74}.  Finally, the ratio of
baryon to meson mass as a function of $N_c$ agrees with the
strong-coupling results of Date, Frishman and Sonnenschein 
\cite{fsxx}.

\begin{figure}
\epsfxsize=3.0in
\centerline{\epsfbox{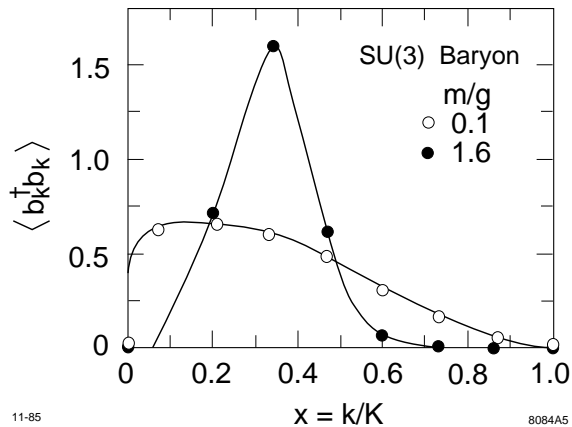}}
\caption[*]{Valence contribution to the baryon structure function in
QCD$_{1+1}$, as a function of the light-cone longitudinal momentum
fraction.  The gauge group is SU(3), $m$ is the quark mass, and $g$
is the gauge coupling \cite{hbp90}.} 
\label{fig1}
\end{figure}

In addition to the spectrum, of course, one obtains the
wavefunctions. These allow direct computation of, \eg, structure
functions.    As an example, Fig. \ref{fig1} shows the valence
contribution to the structure function for an SU(3) baryon, for two
values of the dimensionless coupling $m/g$.  As expected, for weak
coupling the distribution is peaked near $x=1/3$, reflecting that
the baryon momentum is shared essentially equally among its
constituents.  For comparison, the contributions from Fock states
with one and two additional $q\bar{q}$ pairs are shown in Fig. 2.
Note that the amplitudes for these higher Fock components are quite
small relative to the valence configuration.  The lightest hadrons
are nearly always dominated by the valence Fock state in these
super-renormalizable models; higher Fock wavefunctions are typically
suppressed by factors of 100 or more.  Thus the light-cone quarks
are much more like constituent quarks in these theories than
equal-time quarks would be. As discussed above, in an equal-time
formulation even the vacuum state would be an infinite superposition
of Fock states.  Identifying constituents in this case, three of
which could account for most of the structure of a baryon, would be
quite difficult.

\begin{figure}
\centerline{\epsfbox{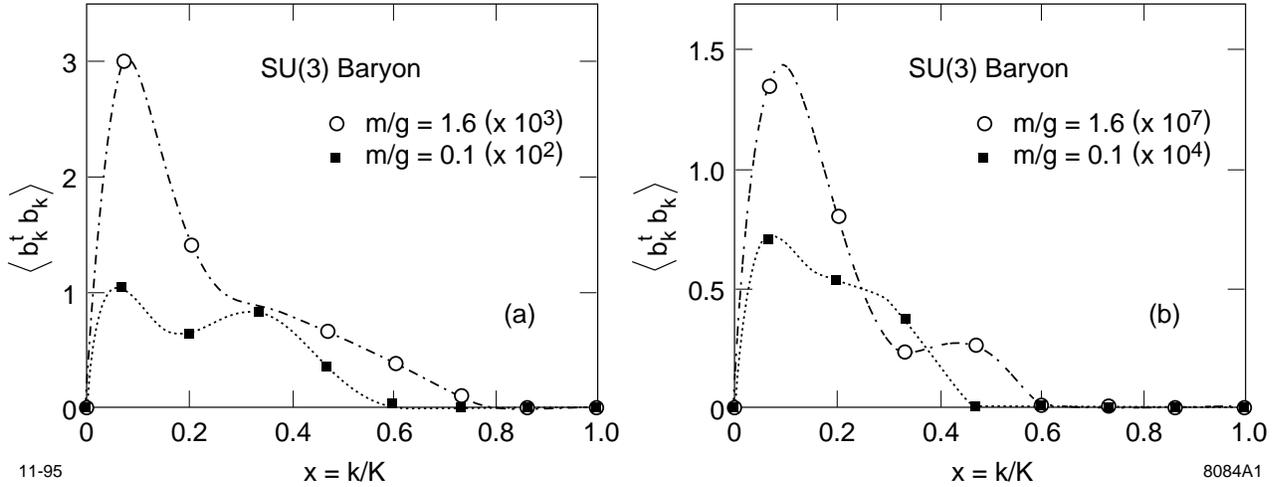}}
\caption[*]{Contributions to the baryon structure function from higher
Fock components: (a) valence plus one additional $q\bar{q}$ pair;
(b) valence plus two additional $q\bar{q}$ pairs \cite{hbp90}.}
\label{fig2}
\end{figure}

\subsection{ Collinear QCD}

QCD can be simplified in a dramatic way by eliminating all
interactions which involve nonzero transverse momentum. The trigluon
interaction is eliminated but the four-gluon and helicity flip $q
\bar q g$ vertices still survive.  In this simplified ``reduced" or
``collinear" theory, one still has all of the degrees of freedom of
QCD(3+1) including transversely polarized color adjoint gluons, but
the theory is effectively a one-space, one-time theory which can be
solved using discretized light-cone quantization.  Recently
Antonuccio and Dalley \cite{Anton} have presented a comprehensive
DLCQ analysis of collinear QCD, obtaining the full physical spectrum
of both quarkonium and gluonium states.  One also obtains the
complete LC Fock wavefunctions for each state of the spectrum.  An
important feature of this analysis is the restoration of complete
rotational symmetry through the degeneracy of states of the rest
frame angular momentum.  In fact as emphasized by Burkardt
\cite{BurkardtParity}, parity and rotational invariance can be
restored if one separately renormalizes the mass that appears in the
helicity-flip $q \bar q g$ vertices and the light-cone  kinetic
energy.

Antonuccio and Dalley \cite{Anton} have also derived ladder
relations which connect the endpoint $x_q \to 0$ behavior of Fock
states with $n$ gluons to the Fock state wavefunction with $n-1$
gluons, relations  which follow most by imposing the condition that
the  $k^+ =0$ mode of the constraint equations vanishes on physical
states. An important condition for a bound state wavefunction is
that gauge invariant quanta have should finite kinetic energy in a
bound state, just as the square of the ``mechanical velocity"
operator $ {\vec v}^2 = (\vec p - e \vec A)^2$ has finite
expectation value in nonrelativistic electrodynamics. Such a
condition automatically connects Fock states of different particle
number.  Thus  the ladder relations should be generalizable to the
full $3 + 1$ theory by requiring that the gauge-extended light-cone
kinetic energy operator have finite expectation value.

\section{EXCLUSIVE PROCESSES AND LIGHT-CONE QUANTIZATION}

A central focus of future QCD studies will be hadron physics at the
amplitude level.  Exclusive reactions such as pion electroproduction
$\gamma^* p \rightarrow n p$ are more subtle to analyze than deep
inelastic lepton scattering and other leading-twist inclusive
reactions since they require the consideration of coherent QCD
effects.  Nevertheless, there is an extraordinary simplification: In
any exclusive reaction where the hadrons are forced to absorb large
momentum transfer $Q$, one can isolate the nonperturbative
long-distance physics associated with hadron structure from the
short-distance quark-gluon hard scattering amplitudes responsible
for the dynamical reaction.  In essence, to leading order in $1/Q,$
each exclusive reaction $AB \rightarrow CD$ factorizes in the form:
\begin{equation}
T_{AB \rightarrow CD}= \int^1_0 \Pi dx_i \phi^\dagger_D(x_i,Q)
\phi^\dagger_C(x_i,Q) \phi_A(x_i,Q) \phi_B(x_i,Q)
T_{\rm quark}\; ,
\end{equation}
where $\phi_A(x_i,Q) = \int^{k_\perp^2 <Q^2} \prod d^2 k_{\perp i}
{\prod d x_i} \psi_{\rm valence}(x_i, k_{\perp i},\lambda_i)$ is the
process-independent distribution amplitude---the light-cone
wavefunction which describes the coupling of hadron $A$ to its
valence quark with longitudinal light-cone momentum fractions $ 0<
x_i < 1$ at impact separation $b= {\cal O} (1/Q)$---and $T_{\rm
quark}$ is the amplitude describing the hard scattering of the
quarks collinear with the hadrons in the initial state to the quarks
which are collinear with the hadrons in the final state.  Since the
propagators and loop momenta in the hard scattering amplitude
$T_{\rm quark}$ are of order $Q$, it can be computed perturbatively
in QCD.  The dimensional counting rules \cite{bf75} for form factors
and fixed CM scattering angle processes follow from the nominal
power-law fall off of $T_{\rm quark}$. The scattering of the quarks
all occurs at short distances; thus the hard scattering amplitude
only couples to the valence-quarks the hadrons when they are at
small relative impact parameter. Remarkably, there are no initial
state or final state interaction corrections to factorization to
leading order in $1/Q$ because of color coherence; final state color
interactions are suppressed.  This feature not only insures the
validity of the factorization theorem for exclusive processes in
QCD, but it also leads to the novel effect of ``color transparency"
in quasi-elastic nuclear reactions \cite{bbgg81,bm88}.

An essential element of the factorization of high momentum transfer
exclusive reactions is universality, \ie, the distribution
amplitudes $\phi_A(x_i,Q)$ are unique wavefunctions specific to each
hadron \cite{bl89}. The distribution amplitudes obey evolution
equations and renormalization group equations \cite{lb80} which can
be derived through the light-cone equations of motion or the
operator product expansion. Thus the same wavefunction that controls
the meson form factors also controls the formation of the mesons in
exclusive decay amplitudes of $B$ mesons such as $B \rightarrow \pi
\pi$ at the comparable momenta.

\section{THE EFFECTIVE CHARGE $\alpha_V(Q^2)$ AND \hfill\break LIGHT-CONE
QUANTIZATION} 

The heavy quark potential plays a central role in QCD, not only in
determining the spectrum and wavefunctions of heavy quarkonium, but
also in providing a physical definition of the running coupling for
QCD.  The heavy quark potential $V(Q^2)$ is defined as the
two-particle irreducible amplitude controlling the scattering of two
infinitely heavy test quarks $Q\overline Q$ in an overall
color-singlet state.  Here $Q^2=-q^2={\vec q}^2$ is the momentum
transfer.  The effective charge $\alpha_V(Q^2)$ is then defined
through the relation $V(Q^2) = - 4 \pi C_F\alpha_V(Q^2)/Q^2$ where
$C_F=(N_c^2-1)/2 N_c = 4/3.$ The running coupling $\alpha_V(Q^2)$
satisfies the usual renormalization group equation, where the first
two terms $\beta_0$ and $\beta_1$ in the perturbation series are
universal coefficients independent of the renormalization scheme or
choice of effective charge.  Thus $\alpha_V$ provides a physical
expansion parameter for perturbative expansions in PQCD.

By definition, all quark and gluon vacuum polarization contributions
are summed into $\alpha_V$; the scale $Q$ of $\alpha_V(Q^2)$ that
appears in perturbative expansions is thus fixed by the requirement
that no terms involving the QCD $\beta$-function appear in the
coefficients.  Thus expansions in $\alpha_V$ are identical to that
of conformally invariant QCD.  This argument is the basis for BLM
scale-fixing \cite{BLM} and commensurate scale relations
\cite{bhjl}, which relate physical observables together without
renormalization scale, renormalization scheme, or other ambiguities
arising from theoretical conventions.

There has recently been remarkable progress \cite{davies} in
determining the running coupling $\alpha_V(Q^2)$ from heavy quark
lattice gauge theory using as input a measured level splitting in
the $\Upsilon$ spectrum.  The heavy quark potential can also be
determined in a direct way from experiment by measuring $e^+ e^- \to
c \bar c$ and $e^+ e^- \to b \bar b$ at threshold \cite{bhkt}.  The
cross section at threshold is strongly modified by the QCD
Sommerfeld rescattering of the heavy quarks through their Coulombic
gluon interactions.  The amplitude near threshold is modified by a
factor $S(\beta,Q^2) = x/(1-\exp(-x))$, where
$x=C_F\alpha_V(Q^2)/\beta$ and $\beta=\sqrt{1-4 m_Q^2)/s}$ is the
relative velocity between the produced quark and heavy quark.  The
scale $Q$ reflects the mean exchanged momentum transfer in the
Coulomb rescattering.  For example, the angular distribution for
$e^+ e^- \to Q\overline Q$ has the form $1 + A(\beta)\cos^2
\theta_{\rm cm}.$ The anisotropy predicted in QCD for small $\beta$
is then $A={\widetilde A}/(1+{\widetilde A})$, where
\begin{equation}
 {\widetilde A}=
      \frac{\beta^2}{2}
      \frac{S(\beta, 4 m_Q^2 \beta^2/e)}{S(\beta, 4 m_Q^2 \beta^2)}
    \frac{1-\frac{4}{\pi} \alpha_V(m_Q^2 \exp 7/6)}{
         1-\frac{16}{3 \pi}\alpha_V(m_Q^2 \exp 3/4)}.
\end{equation}
The last factor is due to hard virtual radiative corrections.  The
anisotropy in $e^+ e^- \to Q \overline Q$ will be reflected in the
angular distribution of the heavy mesons produced in the
corresponding exclusive channels.

The renormalization scheme corresponding to the choice of $\alpha_V$
as the coupling is the natural one for analyzing QCD in the
light-cone formalism, since it automatically sums all vacuum
polarization contributions into the coupling.  For example, once one
knows the form of $\alpha_V(Q^2),$ it can be used directly in the
light-cone formalism as a means to compute the wavefunctions and
spectrum of heavy quark systems.  The effects of the light quarks
and higher Fock state gluons that renormalize the coupling are
already contained in $\alpha_V.$

The same coupling can also be used for computing the hard scattering
amplitudes that control large momentum transfer exclusive reactions
and heavy hadron weak decays.  Thus when evaluating $T_{\rm quark}$
the scale appropriate for each appearance of the running coupling
$\alpha_V$ is the momentum transfer of the corresponding exchanged
gluon \cite{BJiPangLu}.  This prescription agrees with the BLM
procedure.  The connection between $\alpha_V$ and the usual
$\alpha_{\overline{MS}}$ scheme is described elsewhere \cite{bhjl}.

\section{THE PHYSICS OF LIGHT-CONE FOCK\hfill\break STATES}

The light-cone formalism provides the theoretical framework which
allows for a hadron to exist in various Fock configurations.  For
example, quarkonium states not only have valence $Q \overline Q$
components but they also contain $Q\overline Q g$ and $Q \overline Q
g g$ states in which the quark pair is in a color-octet
configuration. Similarly, nuclear LC wave functions contain
components in which the quarks are not in color-singlet nucleon
sub-clusters.  In some processes, such as large momentum transfer
exclusive reactions, only the valence color-singlet Fock state of
the scattering hadrons with small inter-quark impact separation
$b_\perp = {\cal O} (1/Q)$ can couple to the hard scattering
amplitude.  In reactions in which large numbers of particles are
produced, the higher Fock components of the LC wavefunction will be
emphasized. The higher particle number Fock states of a hadron
containing heavy quarks can be diffractively excited, leading to
heavy hadron production in the high momentum fragmentation region of
the projectile. In some cases the projectile's valence quarks can
coalesce with quarks produced in the collision, producing unusual
leading-particle correlations.  Thus the multi-particle nature of
the LC wavefunction can manifest itself in a number of novel ways. 
For example:

\subsection{ Color Transparency }

QCD predicts that the Fock components of a hadron with a small color
dipole moment can pass through nuclear matter without interactions
\cite{bbgg81,bm88}.  Thus in the case of large momentum transfer
reactions, where only small-size valence Fock state configurations
enter the hard scattering amplitude, both the initial and final
state interactions of the hadron states become negligible. 

Color Transparency can be measured though the nuclear dependence of
totally diffractive vector meson production $d\sigma/dt(\gamma^*A\to
V A).$ For large photon virtualities (or for heavy vector
quarkonium), the small color dipole moment of the vector system
implies minimal absorption.  Thus, remarkably, QCD predicts that the
forward amplitude $\gamma^* A \to V A$ at $t \to 0$ is nearly linear
in $A$.  One is also sensitive to corrections from the nonlinear
$A$-dependence of the nearly forward matrix element that couples two
gluons to the nucleus, which is closely related to the nuclear
dependence of the gluon structure function of the nucleus
\cite{bgmfs94}.

The integral of the diffractive cross section over the forward peak
is thus predicted to scale approximately as $A^2/R_A^2 \sim
A^{4/3}.$  Evidence for color transparency in quasi-elastic $\rho$
leptoproduction $\gamma^* A \to \rho^0 N (A-1)$ has recently been
reported by the E665 experiment at Fermilab \cite{e665} for both
nuclear coherent and incoherent reactions. A test could also be
carried out at very small $t_{\rm min}$ at HERA, and would provide a
striking test of QCD in exclusive nuclear reactions. There is also
evidence for QCD ``color transparency" in quasi-elastic $p p$
scattering in nuclei \cite{heppelmann90}.  In contrast to color
transparency, Fock states with large-scale color configurations
interact strongly and with high particle number production
\cite{bbfhs93}.

\subsection{ Hidden Color }

The deuteron form factor at high $Q^2$ is sensitive to wavefunction
configurations where all six quarks overlap within an impact
separation $b_{\perp i} < {\cal O} (1/Q);$ the leading power-law
fall off predicted by QCD is $F_d(Q^2) = f(\alpha_s(Q^2))/(Q^2)^5$,
where, asymptotically, $f(\alpha_s(Q^2))\propto
\alpha_s(Q^2)^{5+2\gamma}$. \cite{bc76}  The derivation of the
evolution equation for the deuteron distribution amplitude and its
leading anomalous dimension $\gamma$ is given elsewhere
\cite{bjl83}. In general, the six-quark wavefunction of a deuteron
is a mixture of five different color-singlet states. The dominant
color configuration at large distances corresponds to the usual
proton-neutron bound state. However at small impact space
separation, all five Fock color-singlet components eventually
acquire equal weight, \ie, the deuteron wavefunction evolves to
80\%\ ``hidden color.''  The relatively large normalization of the
deuteron form factor observed at large $Q^2$ points to sizable
hidden color contributions \cite{fhzxx}.

\subsection{ Spin-Spin Correlations in Nucleon-Nucleon
Scattering and the Charm Threshold }

One of the most striking anomalies in elastic proton-proton
scattering is the large spin correlation $A_{NN}$ observed at large
angles \cite{krisch92}.  At $\sqrt s \simeq 5 $ GeV, the rate for
scattering with incident proton spins parallel and normal to the
scattering plane is four times larger than that for scattering with
anti-parallel polarization. This strong polarization correlation can
be attributed to the onset of charm production in the intermediate
state at this energy \cite{bdt88}.  The intermediate state $\vert u
u d u u d c \bar c \rangle$ has odd intrinsic parity and couples to
the $J=S=1$ initial state, thus strongly enhancing scattering when
the incident projectile and target protons have their spins parallel
and normal to the scattering plane.  The charm threshold can also
explain the anomalous change in color transparency observed at the
same energy in quasi-elastic $ p p$ scattering. A crucial test is
the observation of open charm production near threshold with a cross
section of order of $1 \mu$b.

\subsection{ Anomalous Decays of the $J/\psi$ }

The dominant two-body hadronic decay channel of the $J/\psi$ is
$J/\psi \rightarrow \rho \pi$, even though such vector-pseudoscalar
final states are forbidden in leading order by helicity conservation
in perturbative QCD \cite{tuanxx}.  The $\psi^\prime$, on the other
hand, appears to respect PQCD.  The $J/\psi$ anomaly may signal
mixing with vector gluonia or other exotica \cite{tuanxx}.

\subsection{ The QCD Van Der Waals Potential and
Nuclear Bound Quarkonium }

The simplest manifestation of the nuclear force is the interaction
between two heavy quarkonium states, such as the $\Upsilon (b \bar
b)$ and the $J/\psi(c \bar c)$. Since there are no valence quarks in
common, the dominant color-singlet interaction arises simply from
the exchange of two or more gluons. In principle, one could measure
the interactions of such systems by producing pairs of quarkonia in
high energy hadron collisions. The same fundamental QCD van der
Waals potential also dominates the interactions of heavy quarkonia
with ordinary hadrons and nuclei. The small size of the $Q \overline
Q$ bound state relative to the much larger hadron allows a
systematic expansion of the gluonic potential using the operator
product expansion \cite{manoharxx}.  The coupling of the scalar part
of the interaction to large-size hadrons is rigorously normalized to
the mass of the state via the trace anomaly. This scalar attractive
potential dominates the interactions at low relative velocity. In
this way one establishes that the nuclear force between heavy
quarkonia and ordinary nuclei is attractive and sufficiently strong
to produce nuclear-bound quarkonium \cite{manoharxx,btsxx}.

\subsection{ Anomalous Quarkonium Production at the Tevatron }

Strong discrepancies between conventional QCD predictions and
experiment of a factor of 30 or more have recently been observed for
$\psi$, $\psi^\prime$, and $\Upsilon$ production at large $p_T$ in
high energy $p \overline p$ collisions at the Tevatron \cite{tevxx}.
Braaten and Fleming \cite{bfxx} have suggested that the surplus of
charmonium production is due to the enhanced fragmentation of gluon
jets coupling to the octet $c\overline c$ components in higher Fock
states $\vert c\overline{c}gg\rangle$ of the charmonium
wavefunction. Such Fock states are required for a consistent
treatment of the radiative corrections to the hadronic decay of
$P$-waves in QCD \cite{bblxx}.

\section{INTRINSIC HEAVY QUARK CONTRIBUTIONS IN HADRONIC WAVEFUNCTIONS }

It is important to distinguish two distinct types of quark and gluon
contributions to the nucleon sea measured in deep inelastic
lepton-nucleon scattering: ``extrinsic" and ``intrinsic"
\cite{intcxx}. The extrinsic sea quarks and gluons are created as
part of the lepton-scattering interaction and thus exist over a very
short time $\Delta \tau \sim 1/Q$. These factorizable contributions
can be systematically derived from the QCD hard bremsstrahlung and
pair-production (gluon-splitting) subprocesses characteristic of
leading twist perturbative QCD evolution. In contrast, the intrinsic
sea quarks and gluons are multiconnected to the valence quarks and
exist over a relatively long lifetime within the nucleon bound
state. Thus the intrinsic $q \bar q$ pairs can arrange themselves
together with the valence quarks of the target nucleon into the most
energetically-favored meson-baryon fluctuations.
 
In conventional studies of the ``sea'' quark distributions, it is
usually assumed that, aside from the effects due to
antisymmetrization, the quark and antiquark sea contributions have
the same momentum and helicity distributions. However, the ansatz of
identical quark and antiquark sea contributions has never been
justified, either theoretically or empirically. Obviously the sea
distributions which arise directly from gluon splitting in leading
twist are necessarily CP-invariant; \ie,\ they are symmetric under
quark and antiquark interchange. However, the initial distributions
which provide the boundary conditions for QCD evolution need not be
symmetric since the nucleon state is itself  not CP-invariant. Only
the global quantum numbers of the nucleon must be conserved. The
intrinsic sources of strange (and charm) quarks reflect the
wavefunction structure of the bound state itself; accordingly, such
distributions would not be expected to be CP symmetric. Thus the
strange/anti-strange asymmetry of nucleon structure  functions
provides a direct window into the  quantum bound-state structure of
hadronic wavefunctions.
 
It is also possible to consider the nucleon wavefunction at low
resolution as a fluctuating system coupling  to intermediate
hadronic Fock states such as non-interacting meson-baryon pairs. The
most important fluctuations are most likely to be those closest to
the energy shell and thus have minimal invariant mass.  For example,
the coupling of a proton to a virtual $K^+ \Lambda$ pair provides a
specific source of intrinsic strange quarks and antiquarks in the
proton.   Since the $s$ and $\bar s$ quarks appear in different
configurations in the lowest-lying hadronic pair states, their
helicity and momentum distributions are distinct.
 
Recently Bo-Qiang Ma and I have  investigated the quark and
antiquark asymmetry in the nucleon sea which is implied by a
light-cone meson-baryon fluctuation model of intrinsic $q\bar q$
pairs \cite{BrodMa}. We utilize a boost-invariant light-cone Fock
state description of the hadron wavefunction which emphasizes
multi-parton configurations  of minimal invariant mass. We find that
such fluctuations predict a striking sea quark and antiquark
asymmetry in the corresponding momentum and helicity distributions
in the nucleon structure functions.  In particular, the strange and
anti-strange distributions in the nucleon generally have completely
different momentum and spin characteristics. For example, the model
predicts that the intrinsic $d$ and $s$ quarks in the proton sea are
negatively polarized, whereas the intrinsic $\bar d$ and $\bar s$
antiquarks provide zero contributions to the proton spin. We also
predict that the intrinsic charm and anticharm helicity and momentum
distributions are not strictly identical. We show that the above
picture of quark and antiquark asymmetry in the momentum and
helicity distributions of the nucleon sea quarks has support from a
number of experimental observations, and we suggest processes to
test and measure this quark and antiquark asymmetry in the nucleon
sea.

\subsection {Consequences of Intrinsic Charm and Bottom }

Microscopically, the intrinsic heavy-quark Fock component in the
$\pi^-$ wavefunction, $ \vert \overline u d Q \overline Q \rangle$,
is generated by virtual interactions such as $g g \rightarrow Q
\overline Q$ where the gluons couple to two or more projectile
valence quarks. The probability for $Q \overline Q$ fluctuations to
exist in a light hadron thus scales as $\alpha_s^2(m_Q^2)/m_Q^2$
relative to leading-twist production \cite{vbxx}.  This contribution
is therefore higher twist, and power-law suppressed compared to sea
quark contributions generated by gluon splitting.  When the
projectile scatters in the target, the coherence of the Fock
components is broken and its fluctuations can hadronize, forming new
hadronic systems from the fluctuations \cite{bhmt92}.  For example,
intrinsic $c \overline c$ fluctuations can be liberated provided the
system is probed during the characteristic time $\Delta t = 2p_{\rm
lab}/M^2_{c \overline c}$ that such fluctuations exist. For soft
interactions at momentum scale $\mu$, the intrinsic heavy quark
cross section is suppressed by an additional resolving factor
$\propto \mu^2/m^2_Q$ \cite{chevxx}.  The nuclear dependence arising
from the manifestation of intrinsic charm is expected to be
$\sigma_A\approx \sigma_N A^{2/3}$, characteristic of soft
interactions.

In general, the dominant Fock state configurations are not far off
shell and thus have minimal invariant mass ${\cal M}^2 = \sum_i
m_{T, i}^2/ x_i$ where $m_{T, i}$ is the transverse mass of the
$i^{\rm th}$ particle in the configuration. Intrinsic $Q \overline
Q$ Fock components with minimum invariant mass correspond to
configurations with equal-rapidity constituents. Thus, unlike sea
quarks generated from a single parton, intrinsic heavy quarks tend
to carry a larger fraction of the parent momentum than do the light
quarks \cite{intcxx}.  In fact, if the intrinsic $Q \overline Q$
pair coalesces into a quarkonium state, the momentum of the two
heavy quarks is combined so that the quarkonium state will carry a
significant fraction of the projectile momentum.

There is substantial evidence for the existence of intrinsic $c
\overline c$ fluctuations in the wavefunctions of light hadrons. For
example, the charm structure function of the proton measured by EMC
is significantly larger than that predicted by photon-gluon fusion
at large $x_{Bj}$ \cite{emcicxx}.  Leading charm production in $\pi
N$ and hyperon-$N$ collisions also requires a charm source beyond
leading twist \cite{vbxx,ssnxx}.  The NA3 experiment has also shown
that the single $J/\psi$ cross section at large $x_F$ is greater
than expected from $gg$ and $q \overline q$ production \cite{badxx}.
The nuclear dependence of this forward component is
diffractive-like, as expected from the BHMT mechanism.  In addition,
intrinsic charm may account for the anomalous longitudinal
polarization of the $J/\psi$ at large $x_F$ seen in $\pi N
\rightarrow J/\psi X$ interactions \cite{vanxx}. Further theoretical
work is needed to establish that the data on direct $J/\psi$ and
$\chi_1$ production can be described using the  higher-twist
intrinsic charm mechanism \cite{bhmt92}.

A recent analysis by Harris, Smith and Vogt \cite{hsv95} of the
excessively large charm structure function of the proton at large
$x$ as measured by the EMC collaboration at CERN yields an estimate
that the probability $P_{c \bar c}$ that the proton contains
intrinsic charm Fock states is of the order of $0.6\% \pm 0. 3 \%.$ 
In the case of intrinsic bottom, PQCD scaling predicts
\begin{equation}
P_{ b \bar b}=P_{c \bar c} \frac{m^2_\psi}{m^2_\Upsilon}
\frac{\alpha^4_s(m_b)}{\alpha^4_s(m_c)}\; ,
\end{equation}
more than an order of magnitude smaller.  If super-partners of the
quarks or gluons exist they must also appear in higher Fock states
of the proton, such as $\vert uud ~{\rm gluino}~ {\rm gluino}
\rangle$. At sufficiently high energies, the diffractive excitation
of the proton will produce these intrinsic quarks and gluinos in the
proton fragmentation region.  Such supersymmetric particles can bind
with the valence quarks to produce highly unusual color-singlet
hybrid supersymmetric states such as $\vert uud ~{\rm
gluino}\rangle$ at high $x_F.$ The probability that the proton
contains intrinsic gluinos or squarks scales with the appropriate
color factor and inversely with the heavy particle mass squared
relative to the intrinsic charm and bottom probabilities.  This
probability is directly reflected in the production rate when the
hadron is probed at a hard scale $Q$ which is large compared to the
virtual mass ${\cal M}$ of the Fock state.  At low virtualities, the
rate is suppressed by an extra factor of $Q^2/{\cal M}^2.$ The
forward proton fragmentation regime is a challenge to instrument at
HERA, but it may be feasible to tag special channels involving
neutral hadrons or muons.  In the case of the gas jet fixed-target
$ep$ collisions  such as at HERMES, the target fragments emerge at
low velocity and large backward angles, and thus may be accessible
to precise measurement.

\subsection{ Double Quarkonium Hadroproduction }

It is quite rare for two charmonium states to be produced in the
same hadronic collision.  However, the NA3 collaboration has
measured a double $J/\psi$ production rate significantly above
background in multi-muon events with $\pi^-$ beams at laboratory
momentum 150 and 280 GeV/c and a 400 GeV/c proton beam
\cite{badpxx}.  The relative double to single rate, $\sigma_{\psi
\psi}/\sigma_\psi$, is $(3 \pm 1) \times 10^{-4}$ for pion-induced
production, where $\sigma_\psi$ is the integrated single $\psi$
production cross section.  A particularly surprising feature of the
NA3 $\pi^-N\rightarrow\psi\psi X$ events is that the laboratory
fraction of the projectile momentum carried by the $\psi \psi$ pair
is always very large, $x_{\psi \psi} \geq 0.6$ at 150 GeV/c and
$x_{\psi \psi} \geq 0.4$ at 280 GeV/c.  In some events, nearly all
of the projectile momentum is carried by the $\psi \psi$ system! In
contrast, perturbative $ g g$ and $q \overline q$ fusion processes
are expected to produce central $\psi \psi$ pairs, centered around
the mean value, $\langle x_{\psi\psi} \rangle \approx$ 0.4--0.5, in
the laboratory. There have been attempts to explain the NA3 data
within conventional leading-twist QCD. Charmonium pairs can be
produced by a variety of QCD processes including $B \overline B$
production and decay, $B\overline B \rightarrow \psi \psi X$ and
${\cal O}(\alpha_s^4)$ $\psi \psi$ production via $gg$ fusion and $q
\overline q$ annihilation \cite{esxx,russxx}.  Li and Liu have also
considered the possibility that a $2^{++} c\overline c c \overline
c$ resonance is produced, which then decays into correlated
$\psi\psi$ pairs \cite{llxx}.  All of these models predict centrally
produced $\psi \psi$ pairs \cite{bhkxx,russxx}, in contradiction to
the $\pi^-$ data.

Over a sufficiently short time, the pion can contain Fock states of
arbitrary complexity. For example, two intrinsic $c\overline c$
pairs may appear simultaneously in the quantum fluctuations of the
projectile wavefunction and then, freed in an energetic interaction,
coalesce to form a pair of $\psi$'s.  In the simplest analysis, one
assumes the light-cone Fock state wavefunction is approximately
constant up to the energy denominator \cite{vbxx}. The predicted
$\psi \psi$ pair distributions from the intrinsic charm model
provide a natural explanation of the strong forward production of
double $J/\psi$ hadroproduction, and thus gives strong
phenomenological support for the presence of intrinsic heavy quark
states in hadrons.

It is clearly important for the double $J/\psi$ measurements to be
repeated with higher statistics and at higher energies. The same
intrinsic Fock states will also lead to the production of
multi-charmed baryons in the proton fragmentation region. The
intrinsic heavy quark model can also be used to predict the features
of heavier quarkonium hadroproduction, such as $\Upsilon \Upsilon$,
$\Upsilon \psi$, and $(c\bar b)$ $(\bar cb)$ pairs. It is also
interesting to study the correlations of the heavy quarkonium pairs
to search for possible new four-quark bound states and final state
interactions generated by multiple gluon exchange \cite{llxx}, 
\hbox{since}
the QCD Van der Waals interactions could be anomalously strong at
low relative rapidity \cite{manoharxx,btsxx}.

\subsection{ Leading Particle Effect in Open Charm Production }

According to PQCD factorization, the fragmentation of a heavy quark
jet is independent of the production process. However, there are
strong correlations between the quantum numbers of $D$ mesons and
the charge of the incident pion beam in $\pi N \rightarrow D X$
reactions. This effect can be explained as being due to the
coalescence of the produced intrinsic charm quark with co-moving
valence quarks. The same higher-twist recombination effect can also
account for the suppression of $J/\psi$ and $\Upsilon$ production in
nuclear collisions in regions of phase space with high particle
density \cite{vbxx}.

There are other ways in which the intrinsic heavy quark content of
light hadrons can be tested. More measurements of the charm and
bottom structure functions at large $x_F$ are needed to confirm the
EMC data \cite{emcicxx}.  Charm production in the proton
fragmentation region in deep inelastic lepton-proton scattering is
sensitive to the hidden charm in the proton wavefunction. The
presence of intrinsic heavy quarks in the hadron wavefunction also
enhances heavy flavor production in hadronic interactions near
threshold.  More generally, the intrinsic heavy quark model leads to
enhanced open and hidden heavy quark production and leading particle
correlations at high $x_F$ in hadron collisions, with a distinctive
strongly shadowed nuclear dependence characteristic of soft hadronic
collisions.

It is of particular interest to examine the fragmentation of the
proton when the electron strikes a light quark and the interacting
Fock component is the $\vert uud c \bar c \rangle$ or $\vert uud b
\bar b \rangle$ state.  These Fock components correspond to
intrinsic charm or intrinsic bottom quarks in the proton
wavefunction.  Since the heavy quarks in the proton bound state have
roughly the same rapidity as the proton itself, the intrinsic heavy
quarks will appear at large $x_F$.  One expects heavy quarkonium and
also heavy hadrons to be formed from the coalescence of the heavy
quark with the valence $u$ and $d$ quarks, since they have nearly
the same rapidity.  Since the heavy and valence quark momenta
combine, these states are preferentially produced with large
longitudinal momentum fractions

The role of intrinsic charm becomes dominant over leading-twist
fusion processes near threshold, since the multi-connected intrinsic
charm configurations in the higher light-cone Fock state of the
proton are more efficient that gluon splitting in producing charm.
The heavy $c$ and $\bar c$ will be produced at low velocities
relative to each other and with the spectator quarks from the proton
and virtual photon.  As is the case of $e^+ e^- \to \bar c c$ near
threshold, the QCD Coulomb rescattering will give Sommerfeld
correction factors $S(\beta,Q^2)$ which strongly distort the Born
predictions for the production amplitudes.

\section{THE FORM FACTORS OF ELEMENTARY AND COMPOSITE SYSTEMS}

In this section I will review the light-cone formalism for both
elementary and composite systems \cite{r12,r13,bd80}.  We choose
light-cone coordinates with the incident lepton directed along the
$z$ direction \cite{r14} $(p^\pm\equiv p^0\pm p^3)$:
\begin{equation}
p^\mu \equiv (p^+,p^-,\longvec p_1) =
\left(p^+,\frac{M^2}{p^+},\longvec 0\!\!_\perp\right)\ ,\qquad
q = \left(0,\frac{2q\cdot p}{p^+}, \longvec q\!\!_\perp\right)\ ,
\label{eq6}
\end{equation}
where $q^2=-2q\cdot p=-q^2\!\!_\perp$ and $M=m_\ell$ is the mass of
the composite system.  The Dirac and Pauli form factors can be
identified \cite{r13} from the spin-conserving and spin-flip current
matrix elements $(J^+=J^0+J^3)$:
\begin{eqnarray}
{\cal M}^+_{\uparrow\uparrow} &=&
\left\langle p+q,\uparrow\left|\frac{J^+(0)}{p^+}\right|p,\uparrow
\right\rangle = 2F_1(q^2) \ ,
\label{eq7}\\[5pt]
{\cal M}^+_{\uparrow\downarrow} &=&
\left\langle p+q,\uparrow\left|\frac{J^+(0)}{p^+}\right|p,\uparrow
\right\rangle = - 2(q_1-iq_2) \ \frac{F_2(q^2)}{2M} \ ,
\label{eq8}
\end{eqnarray}
where $\uparrow$ corresponds to positive spin projection $S_z=+
\frac{1}{2}$ along the $\widehat z$ axis.

Each Fock-state wave function $\left|n\right\rangle$ of the incident
lepton is represented by the functions $\psi^{(n)}_{p,S_z}(x_i,\longvec
k\!\!_{\perp i},S_i)$, where
\[ k^\mu \equiv (k^+,k^-,\longvec k\!\!_\perp) = \left( xp^+,\,
\frac{k^2_\perp+m^2}{xp^+},\longvec k\!\!_\perp\right)
\]
specifies the light-cone momentum coordinates of each constituent
$i=1,\ldots,n$, and $S_i$ specifies its spin projection $S^i_z$.
Momentum observation on the light cone requires
\[ \sum^n_{i=1} k_{\perp i} = 0 \ , \qquad
\sum^n_{i=1} x_i= 1 \ ,
\]
and thus $0<x_i<1$.  The amplitude to find $n$ (on-mass-shell)
constituents in the lepton is then $\psi^{(n)}$ multiplied by the
spinor factors $u_{S_i}(k_i)/(k^+_i)^{-1/2}$ or
$v_{S_i}(k_i)/(k^+_i)^{1/2}$ for each constituent fermion or
anti-fermion \cite{r15}.  The Fock state is off the ``energy shell'':
\[ \left(p^--\sum^n_{i=1}k^-_i\right) p^+ = \sum^n_{i=1}
\left(\frac{\longvec k^2\!\!_{\perp i}+m^2_i}{x_i}\right) \ .
\]
The quantity $(\longvec k^2\!\!_{\perp i}+m^2_i)/x_i$ is the
relativistic analog of the kinetic energy $\longvec p^2_i/2m_i$ in
the Schr\"odinger formalism.

The wave function for the lepton directed along the final direction
$p+q$ in the current matrix element is then
\[ \psi^{(n)}_{p+q,S^\prime_z} (x_i,\longvec k^\prime\!\!_{\perp i},
S^\prime_i) \ ,
\]
where \cite{dy70}
\[ \longvec k^\prime\!\!_{\perp j}= \longvec k\!\!_{\perp j}
+(1-x_j)\longvec q\!\!_\perp
\]
for the struck constituent and
\[ \longvec k^\prime\!\!_{\perp i} = \longvec k\!\!_{\perp i} -
x_i\longvec q\!\!_\perp
\]
for each spectator $(i\ne j)$.  The $\longvec k{}^\prime\!\!_\perp$
are transverse to the $p+q$ direction with
\[ \sum^n_{i=1} \longvec k^\prime\!\!_{\perp i} = 0 \ .
\]

The interaction of the current $J^+(0)$ conserves the spin
projection of the struck constituent fermion $(\bar
u_s,\gamma^+u_s)/k_+=2\delta_{ss^\prime}$.  Thus from Eqs.
(\ref{eq7}) and (\ref{eq8})
\begin{equation}
F_1(q^2) = \frac{1}{2} {\cal M}^+_{\uparrow\uparrow} =
\sum_j e_j \int [dx]\,\left[d^2\longvec k\!\!_\perp\right]\,
\psi^{*(n)}_{p+q,\uparrow}\left(x,\longvec
k^\prime\!\!_\perp,S\right)\,
\psi^{(n)}_{p,\uparrow}\left(x,\longvec k\!\!_\perp,S\right) \ ,
\label{eq9} 
\end{equation}
and
\begin{eqnarray}
\lefteqn{-\left(\frac{q_1-iq_2}{2M}\right) F_1(q^2)  = 
\frac{1}{2} {\cal M}^+_{\uparrow\downarrow}\nonumber}
\hspace{30pt} \\[5pt] &=&
\sum_j e_j \int [dx]\,\left[d^2\longvec k\!\!_\perp\right]\,
\psi^{*(n)}_{p+q,\uparrow}\left(x,\longvec
k^\prime\!\!_\perp,S\right)\,
\psi^{(n)}_{p,\uparrow}\left(x,\longvec k\!\!_\perp,S\right) \ ,
\label{eq10}
\end{eqnarray}
where $e_j$ is the fractional charge of each constituent.  [A
summation of all possible Fock states $(n)$ and spins $(S)$ is
assumed.]  The phase-space integration is
\begin{equation}
[dx] \equiv \delta \left(1-\sum x_i\right) \prod^n_{i=1} dx_i \ ,
\label{eq11}
\end{equation}
and
\begin{equation}
\left[ d^2k_\perp\right] \equiv 16\pi^3\delta^{(2)}
\left(\sum k_{\perp i}\right) \prod^n_{i=1}
\frac{d^2k_\perp}{16\pi^3} \ .
\label{eq12}
\end{equation}
Equation (\ref{eq9}) evaluated at $q^2=0$ with $F_1(0)=1$ is
equivalent to wave-function normalization.  The anomalous moment
$a=F_2(0)/F_1(0)$ can be determined from the coefficient linear in
$q_1-iq_2$ from the coefficient linear in $q_1-iq_2$ from
$\psi^*_{p+q}$ in Eq. (\ref{eq10}).  In fact, since \cite{r17}
\begin{equation}
\frac{\partial}{\partial\longvec q\!\!_\perp}\, \psi^*_{p+q}
\equiv - \sum_{i\ne j}x_i \frac{\partial}{\partial\longvec
k\!\!_{\perp i}}\, \psi^*_{p+q}
\label{eq13}
\end{equation}
(summed over spectators), we can, after integration by parts, write
explicitly
\begin{equation}
\frac{a}{M} = - \sum_je_j\int [dx]\int \left[d^2k_\perp\right]
\sum_{i\ne j} \psi^*_{p \uparrow}x_i \left(\frac{\partial}{\partial
k_{1i}}+i\frac{\partial}{\partial k_{2i}}\right) \psi_{p\downarrow} \
. 
\label{eq14}
\end{equation}
The wave function normalization is
\begin{equation}
\int [dx]\int \left[d^2k_\perp\right]\, \psi^*_{p\uparrow}\,
\psi_{p\uparrow} = \int [dx]\int d^2k_\perp\psi^*_{p\downarrow}\,
\psi_{p\downarrow} = 1 \ .
\label{eq15}
\end{equation}
A sum over all contributing Fock states is assumed in Eqs.
(\ref{eq14}) and (\ref{eq15}).

We thus can express the anomalous moment in terms of a local matrix
element at zero momentum transfer.  It should be emphasized that Eq.
(\ref{eq14}) is exact; it is valid for the anomalous element of any
spin-$\frac{1}{2}$ system.

As an example, in the case of the electron's anomalous moment to
order $\alpha$ in QED, \cite{r18} the contributing intermediate Fock
states  are the electron-photon states with spins
$\left| -\frac{1}{2},1\right\rangle$ and
$\left|\frac{1}{2},-1\right\rangle$:
\begin{equation}
\psi_{p\downarrow} = \frac{e/\sqrt x}
{M^2-\frac{k^2_\perp+\lambda^2}{x}-\frac{k^2_\perp+\widehat m^2}{1-x}}
\times\left\{\begin{array}{l}
\sqrt 2\  \frac{(k_1-ik_2)}{x}\
\left(\left|-\frac{1}{2}\right\rangle\rightarrow\left|-\frac{1}{2},1
\right \rangle\right) \\[5pt]
\sqrt 2\  \frac{M(1-x)-\widehat m}{1-x}\
\left(\left|-\frac{1}{2}\right\rangle\rightarrow\left|\frac{1}{2},-1
\right\rangle\right)
\end{array} \right.
\label{eq16}
\end{equation}
and
\begin{equation}
\psi^*_{p\uparrow} = \frac{e/\sqrt x}
{M^2-\frac{k^2_\perp+\lambda^2}{x}-\frac{k^2_\perp+
\widehat m^2}{1-x}} \times\left\{
\begin{array}{l}
-\sqrt 2\  \frac{M(1-x)-\widehat m}{1-x}\
\left(\left|-\frac{1}{2},1\right\rangle\rightarrow\left|
\frac{1}{2}\right\rangle\right)\\[5pt]
-\sqrt 2\  \frac{(k_1-ik_2)}{x}\
\left(\left|\frac{1}{2},-1\right\rangle
\rightarrow \left|\frac{1}{2}\right\rangle\right) \ .
\end{array} \right.
\label{eq17}
\end{equation}
The quantities to the left of the curly bracket in Eqs. (\ref{eq16})
and (\ref{eq17}) are the matrix elements of
\[ \frac{\bar u}{(p^+-k^+)^{1/2}}\, \gamma\cdot\epsilon^*\,
\frac{u}{(p^+)^{1/2}}\quad \hbox{and}\quad
\frac{\bar u}{(p^+)^{1/2}}\, \gamma\cdot\epsilon\,
\frac{u}{(p^+-k^+)^{1/2}} \ ,
\]
respectively, where $\widehat \epsilon =
\widehat\epsilon_{\uparrow(\downarrow)}=\pm (1/\sqrt 2)(\widehat
x\pm i\widehat y)$, $\epsilon\cdot k=0$, $\epsilon^+=0$ in the
light-cone gauge for vector spin projection $S_z=\pm 1$
\cite{r12,r13}. For the sake of generality, we let the intermediate
lepton and vector boson have mass $\widehat m$ and $\lambda$,
respectively.

Substituting (\ref{eq16}) and (\ref{eq17}) into Eq. (\ref{eq14}),
one finds that only the $\left|-\frac{1}{2},1\right\rangle$
intermediate state actually contributes to $a$, since terms which
involve differentiation of the denominator of $\psi_{p\downarrow}$
cancel.  We thus have \cite{bd80}
\begin{equation}
a = 4M\, e^2\int \frac{d^2k_\perp}{16\pi^3}\int^1_0dx\
\frac{\left[\widehat
m-(1-x)M\right]/x(1-x)}{\left[M^2-(k^2_\perp+\widehat
m^2)/(1-x)-(k^2_\perp+\lambda^2)/x\right]^2} \ ,
\label{eq18}
\end{equation}
or
\begin{equation}
a = \frac{\alpha}{\pi}\int^1_0dx\
\frac{M\left[\widehat m-M(1-x)\right]x(1-x)}{\widehat
m^2x+\lambda^2(1-x)-M^2x(1-x)} \ ,
\label{eq19}
\end{equation}
which, in the case of QED $(\widehat m=M, \lambda=0)$ gives the
Schwinger results $a=\alpha/2\pi$.

The general result (\ref{eq14}) can also be written in matrix form:
\begin{equation}
\frac{a}{2M} = - \sum_j e_j \int [dx] \left[d^2k_\perp\right]\,
\psi^+\longvec S\!\!_\perp\cdot\longvec L\!\!_\perp\psi \ ,
\label{eq20}
\end{equation}
where $S$ is the spin operator for the total system and $\longvec
L\!\!_\perp$ is the generator of ``Galiean'' transverse boosts
\cite{r12,r13} on the light cone, \sl \ie, \rm $\longvec
S\!\!_\perp\cdot \longvec L\!\!_\perp = (S_+L_-+S_-L_+)/2$ where
$S_\pm = (S_1\pm iS_2)$ is the spin-ladder operator and
\begin{equation}
L_\pm = \sum_{i\ne j} x_i \left(\frac{\partial}{\partial k_{\perp
i}} \mp i\, \frac{\partial}{\partial k_{2i}}\right)
\label{eq21}
\end{equation}
(summed over spectators) in the analog of the angular momentum
operator $\longvec p\times \longvec r$.  Equation (\ref{eq14}) can
also be written simply as an expectation value in impact space.

The results given in Eqs. (\ref{eq9}), (\ref{eq10}), and
(\ref{eq14}) may also be convenient for calculating the anomalous
moments and form factors of hadrons in quantum chromodynamics
directly from the quark and gluon wave functions $\psi(\longvec
k\!\!_\perp, x, S)$.  These wave functions can also be used to
construct the structure functions and distribution amplitudes which
control large momentum transfer inclusive and exclusive processes
\cite {r13,r19}.  The charge radius of a composite system can also
be written in the form of a local, forward matrix element \cite{r20}:
\begin{equation}
\frac{\partial F_1(q^2)}{\partial q^2}\Bigg|_{q^2=0} = - \sum _j
e_j \int [dx]\, \left[d^2k_\perp\right]\, \psi^*_{p,\uparrow}
\left(\sum_{i\ne j} x_i\, \frac{\partial}{\partial\longvec
k\!\!_{\perp i}}\right)^2 \psi_{p,\uparrow} \ .
\label{eq22}
\end{equation}

We thus find that, in general, any Fock state $\left|n\right\rangle$
which couples to both $\psi^*_\uparrow$ and $\psi_\downarrow$ will
give a contribution to the anomalous moment.  Notice that because of
rotational symmetry in the $\widehat x, \widehat y$ direction, the
contribution to $a=F_2(0)$ in Eq. (\ref{eq14}) always involves the
form $(a,b=1,\ldots,n)$
\begin{equation}
M\psi^*_\uparrow \sum_{i\ne j} x_i\, \frac{\partial}{\partial k_{\perp
i}}\, \psi_\downarrow \sim \mu M \rho \left(\longvec
k^a\!\!_\perp\cdot \longvec k^b\!\!_\perp\right) \ ,
\label{eq23}
\end{equation}
compared to the integral (\ref{eq15}) for wave-function
normalization which has terms of order
\[ \psi^*_\uparrow \psi_\uparrow \sim \longvec k^a\!\!_\perp \cdot
\longvec k^b\!\!_\perp \rho \left(\longvec k^a\!\!_\perp\cdot
\longvec k^b\!\!_\perp\right)
\]
and
\begin{equation}
\mu^2 \rho \left(\longvec k^a\!\!_\perp\cdot k^b_\perp\right) \ .
\label{eq24}
\end{equation}
here $\rho$ is a rotationally invariant function of the transverse
momenta and $\mu$ is a constant with dimensions of mass.  Thus, in
order of magnitude
\begin{equation}
a = {\cal O}\, \left(\frac{\mu M}{\mu^2+\left\langle\longvec
k^2\!\!_\perp\right\rangle}\right)
\label{eq25}
\end{equation}
summed and weighted over the Fock states.  In the case of a
renormalizable theory, the only parameters $\mu$ with the dimension
of mass are fermion masses.  In super-renormalizable theories, $\mu$
can be proportional to a coupling constant $g$ with dimension of
mass \cite{r21}.

In the case where all the mass-scale parameters of the composite
state are of the same order of magnitude, we obtain $a={\cal O}(MR)$
as in Eqs. (\ref{eq11}) and (\ref{eq12}), where $R= \left\langle
k^2_\perp\right\rangle^{-1/2}$ is the characteristic size \cite{r22}
of the Fock State.  On the other hand, in theories where $\mu^2 \ll
\left\langle k^2_\perp\right\rangle$, we obtain the quadratic
relation $a ={\cal O}(\mu\, MR^2)$.

Thus composite models for leptons can avoid conflict with the
high-precision QED measurements in several ways.
\begin{itemize}
\item
There can be strong cancellations between the contribution of
different Fock \hbox{states}.  
\item
The parameter $\mu$ can be minimized.  For example, in a
renormalizable theory this can be accomplished by having the bound
state of light fermions and heavy bosons.  Since $\mu \geq M$, we
then have $a \geq {\cal O}(M^2R^2)$.
\item
If the parameter $\mu$ is of the same order s the other mass scales
in the composite state, then we have a linear condition $a = {\cal
O}(MR)$. 
\end{itemize}

\section{MOMENTS OF NUCLEONS AND NUCLEI IN THE LIGHT-CONE FORMALISM}

The use of covariant kinematics leads to a number of striking
conclusions for the electromagnetic and weak moments of nucleons and
nuclei. For example, magnetic moments cannot be written as the naive
sum $\overrightarrow\mu = \sum\overrightarrow\mu_i$ of the magnetic
moments of the constituents, except in the nonrelativistic limit
where the radius of the bound state is much larger than its Compton
scale: $R_A M_A\gg 1$. The deuteron quadrupole moment is in general
nonzero even if the nucleon-nucleon bound state has no $D$-wave
component \cite{bh83}.  Such effects are due to the fact that even
``static'' moments must be computed as transitions between states of
different momentum $p^\mu$ and $p^\mu + q^\mu$, with $q^\mu
\rightarrow 0$. Thus one must construct current matrix elements
between boosted states. The Wigner boost generates nontrivial
corrections to the current interactions of bound systems
\cite{bp69}.  Remarkably, in the case of the deuteron, both the
quadrupole and magnetic moments become equal to that of the Standard
Model in the limit $M_d R_d\rightarrow 0.$ In this limit, the three
form factors of the deuteron have the same ratios as do those of the
$W$ boson in the Standard Model \cite{bh83}.

One can also use light-cone methods to show that the proton's
magnetic moment $\mu_p$ and its axial-vector coupling $g_A$ have a
relationship independent of the specific form of the light-cone
wavefunction \cite{bs94}.  At the physical value of the proton
radius computed from the slope of the Dirac form factor, $R_1=0.76$
fm, one obtains the experimental values for both $\mu_p$ and $g_A$;
the helicity carried by the valence $u$ and $d$ quarks are each
reduced by a factor $\simeq 0.75$ relative to their nonrelativistic
values. At infinitely small radius $R_p M_p\rightarrow 0$, $\mu_p$
becomes equal to the Dirac moment, as demanded by the
Drell-Hearn-Gerasimov sum rule \cite{gerasimov65,dh66}.  Another
surprising fact is that as $R_1 \rightarrow 0$ the constituent quark
helicities become completely disoriented and $g_A \rightarrow 0$.

One can understand the origins of the above universal features even
in an effective three-quark light-cone Fock description of the
nucleon. In such a model, one assumes that additional degrees of
freedom (including zero modes) can be parameterized through an
effective potential \cite{lb80}.  After truncation, one could in
principle obtain the mass $M$ and light-cone wavefunction of the
three-quark bound-states by solving the Hamiltonian eigenvalue
problem. It is reasonable to assume that adding more quark and
gluonic excitations will only refine this initial approximation
\cite{phw90}. In such a theory the constituent quarks will also
acquire effective masses and form factors.

Since we do not have an explicit representation for the effective
potential in the light-cone Hamiltonian $P^-_{\rm eff}$ for three
quarks, we shall proceed by making an Ansatz for the momentum-space
structure of the wavefunction $\Psi$.  Even without explicit
solutions of the Hamiltonian eigenvalue problem, one knows that the
helicity and flavor structure of the baryon eigenfunctions will
reflect the assumed global SU(6) symmetry and Lorentz invariance of
the theory.  As we will show below, for a given size of the proton
the predictions and interrelations between observables at $Q^2=0,$
such as the proton magnetic moment $\mu_p$ and its axial coupling
$g_A,$ turn out to be essentially independent of the shape of the
wavefunction \cite{bs94}.

The light-cone model given by Ma \cite{Ma91}  and by Schlumpf
\cite{schlumpf93} provides a framework for representing the general
structure of the effective three-quark wavefunctions for baryons.
The wavefunction $\Psi$ is constructed as the product of a momentum
wavefunction, which is spherically symmetric and invariant under
permutations, and a spin-isospin wave function, which is uniquely
determined by SU(6)-symmetry requirements.  A Wigner-Melosh rotation
\cite{wigner39,melosh74} is applied to the spinors, so that the
wavefunction of the proton is an eigenfunction of $J$ and $J_z$ in
its rest frame \cite{strik,coester82,ls78}.  To represent the range
of uncertainty in the possible form of the momentum wavefunction,
one can choose two simple functions of the invariant mass ${\cal M}$
of the quarks:
\begin{eqnarray} 
\psi_{\rm H.O.}({\cal M}^2) &=& N_{\rm H.O.}\exp(-{\cal
M}^2/2\beta^2),\\ \psi_{\rm Power}({\cal M}^2) &=& N_{\rm Power}
(1+{\cal M}^2/\beta^2)^{-p}\; ,
\end{eqnarray} 
where $\beta$ sets the characteristic internal momentum scale.
Perturbative QCD predicts a nominal power-law fall off at large
$k_\perp$ corresponding to $p=3.5$ \cite{lb80}.  The Melosh rotation
insures that the nucleon has $j=\frac{1}{2}$ in its rest system.  It
has the matrix representation \cite{melosh74}
\begin{equation}
R_M(x_i,k_{\perp i},m)=\frac{m+x_i {\cal M}-i\overrightarrow
\sigma\cdot(\vec n\times\vec k_i)}{\sqrt{(m+x_i {\cal M})^2+
k_{\perp i}^2} } 
\end{equation} 
with $\vec n=(0,0,1)$, and it becomes the unit matrix if the quarks
are collinear, $R_M(x_i,0,m)=1.$ Thus the internal transverse
momentum dependence of the light-cone wavefunctions also affects its
helicity structure \cite{bp69}.

The Dirac and Pauli form factors $F_1(Q^2)$ and $F_2(Q^2)$ of the
nucleons are given by the spin-conserving and the spin-flip matrix
elements of the vector current $J^+_V$ (at $Q^2=-q^2$) \cite{bd80}
\begin{eqnarray}
F_1(Q^2) &=& \langle p+q,\uparrow | J^+_V |
p,\uparrow \rangle , \\
(Q_1-i Q_2) F_2(Q^2) &=& -2M\langle
p+q,\uparrow | J^+_V | p, \downarrow \rangle \; .
\end{eqnarray}
We then can calculate the anomalous magnetic moment
$a=\lim_{Q^2\rightarrow 0} F_2(Q^2)$.\footnote{The total proton
magnetic moment is $\mu_p = \frac{e}{2M}(1+a_p).$} The same
parameters as given by Schlumpf \cite{schlumpf93} are chosen, namely
$m=0.263$ GeV (0.26 GeV) for the up (down) quark masses,
$\beta=0.607$ GeV (0.55 GeV) for $\psi_{\rm Power}$ ($\psi_{\rm
H.O.}$), and $p=3.5$. The quark currents are taken as elementary
currents with Dirac moments $\frac{e_q}{2 m_q}.$ All of the baryon
moments are well-fit if one takes the strange quark mass as 0.38
GeV. With the above values, the proton magnetic moment is 2.81
nuclear magnetons, and the neutron magnetic moment is $-1.66$
nuclear magnetons. (The neutron value can be improved by relaxing
the assumption of isospin symmetry.) The radius of the proton is
0.76 fm, \ie, $M_p R_1=3.63$.

\begin{figure}
\centerline{\epsfbox{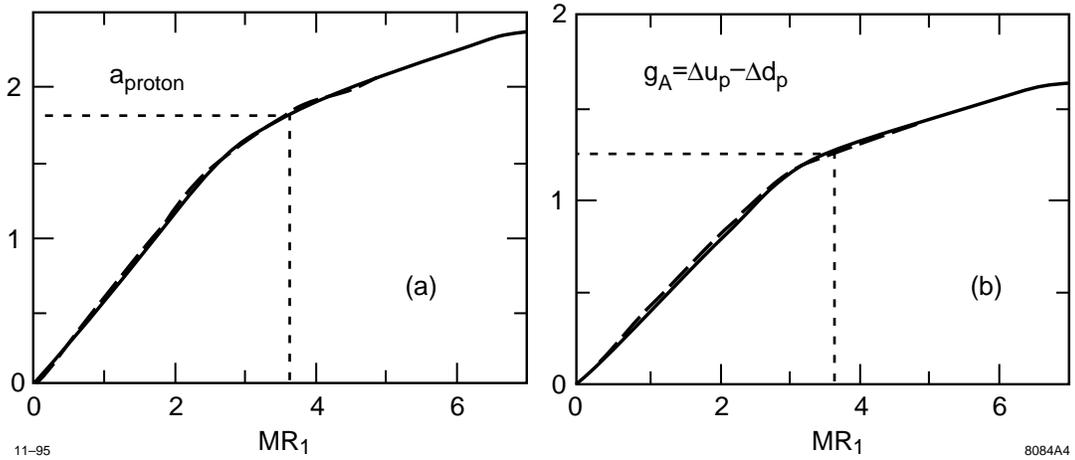}}
\caption[*]{(a).
The anomalous magnetic moment of the proton $a_p=F_2(0)$ as  a
function of its Dirac radius $M_p R_1 $ in Compton units. (b). The
axial vector coupling of the neutron to proton beta-decay as a
function of $M_p R_1.$ In each figure, the broken line is computed
from a wavefunction with power-law fall off  and the solid curve is
computed from a Gaussian wavefunction. The experimental values at
the physical proton Dirac radius are indicated by the dotted line
\cite {bs94}.}
\label{fig3}
\end{figure}

In Fig.~\ref{fig3}(a) we show the functional relationship between
the anomalous moment $a_p$ and its Dirac radius predicted by the
three-quark light-cone model. The value of
\begin{equation}
R^2_1 = -6 \frac{dF_1(Q^2)}{dQ^2}\Bigl\vert_{Q^2=0}
\end{equation}
is varied by changing $\beta$ in the light-cone wavefunction while
keeping the quark mass $m$ fixed.  The prediction for the power-law
wavefunction $\psi_{\rm Power}$ is given by the broken line; the
continuous line represents $\psi_{\rm H.O.}$.  Figure~\ref{fig3}(a)
shows that when one plots the dimensionless observable $a_p$ against
the dimensionless observable $M R_1$ the prediction is essentially
independent of the assumed power-law or Gaussian form of the
three-quark light-cone wavefunction.  Different values of $p>2$ also
do not affect the functional dependence of $a_p(M_p R_1)$ shown in
Fig.~\ref{fig3}(a). In this sense the predictions of the three-quark
light-cone model relating the $Q^2 \rightarrow 0$ observables are
essentially model-independent. The only parameter controlling the
relation between the dimensionless observables in the light-cone
three-quark model is $m/M_p$ which is set to 0.28. For the physical
proton radius $M_p R_1=3.63$ one obtains the empirical value for
$a_p=1.79$ (indicated by the dotted lines in Fig. \ref{fig3}(a)).

\setcounter{footnote}{0}

The prediction for the anomalous moment $a$ can be written
analytically as $a=\langle \gamma_V \rangle a^{\rm NR}$, where
$a^{\rm NR}=2M_p/3m$ is the nonrelativistic ($R\rightarrow\infty$)
value and $\gamma_V$ is given as \cite{cc91}
\begin{equation}
\gamma_V(x_i,k_{\perp i},m)= \frac{3m}{{\cal M}}
\left[\frac{(1-x_3){\cal M}(m+x_3 {\cal M})- \vec k_{\perp 3}^2/2}
{(m+x_3 {\cal M})^2+\vec k_{\perp 3}^2}\right]\; .
\end{equation}
The expectation value $\langle \gamma_V \rangle$ is evaluated
as\footnote{Here $[d^3k]\equiv d\vec k_1d\vec k_2d\vec
k_3\delta(\vec k_1+\vec k_2+ \vec k_3)$. The third component of
$\vec k$ is defined as $k_{3i}\equiv\frac{1}{2}(x_i{\cal M}-\frac{m^2+\vec
k_{\perp i}^2}{x_i {\cal M}})$. This measure differs from the
usual one used \cite{lb80} by the Jacobian $\prod
\frac{dk_{3i}}{dx_i}$ which can be absorbed into the wavefunction.}
\begin{equation} 
\langle\gamma_V\rangle = \frac{\int [d^3k] \gamma_V |\psi|^2}{\int
[d^3k] |\psi|^2}\; . 
\end{equation}

Let us now take a closer look at the two limits $R \rightarrow
\infty$ and $R\rightarrow 0$. In the nonrelativistic limit we let
$\beta \rightarrow 0$ and keep the quark mass $m$ and the proton
mass $M_p$ fixed. In this limit the proton radius $R_1 \rightarrow
\infty$ and $a_p \rightarrow 2M_p/3m = 2.38$, since $\langle
\gamma_V \rangle \rightarrow 1$.\footnote{This differs slightly from
the usual nonrelativistic formula $1+a=\sum_q \frac{e_q}{e}
\frac{M_p}{
m_q}$ due to the non-vanishing binding energy which results in $M_p
\neq 3m_q$.} Thus the physical value of the anomalous magnetic
moment at the empirical proton radius $M_p R_1=3.63$ is reduced by
25\% from its nonrelativistic value due to relativistic recoil and
nonzero $k_\perp$.\footnote{The nonrelativistic value of the neutron
magnetic moment is reduced by 31\%.}

To obtain the ultra-relativistic limit we let $\beta \rightarrow
\infty$ while keeping $m$ fixed.  In this limit the proton becomes
pointlike, $M_p R_1 \rightarrow 0$, and the internal transverse
momenta $k_\perp \rightarrow\infty$. The anomalous magnetic momentum
of the proton goes linearly to zero as $a=0.43 M_p R_1$ since
$\langle\gamma_V\rangle\rightarrow 0$.  Indeed, the
Drell-Hearn-Gerasimov sum rule \cite{gerasimov65,dh66} demands that
the proton magnetic moment become equal to the Dirac moment at small
radius.  For a spin-$\frac{1}{2}$ system
\begin{equation}
a^2=\frac{M^2}{2\pi^2\alpha}\int_{s_{th}}^\infty \frac{ds}{s}\left[
\sigma_P(s)-\sigma_A(s)\right]\; ,
\end{equation}
where $\sigma_{P(A)}$ is the total photoabsorption cross section
with parallel (anti-parallel) photon and target spins. If we take
the point-like limit, such that the threshold for inelastic
excitation becomes infinite while the mass of the system is kept
finite, the integral over the photoabsorption cross section vanishes
and $a=0$. \cite{bd80}  In contrast, the anomalous magnetic moment
of the proton does not vanish in the nonrelativistic quark model as
$R\rightarrow 0$. The nonrelativistic quark model does not reflect
the fact that the magnetic moment of a baryon is derived from lepton
scattering at nonzero momentum transfer, \ie, the calculation of a
magnetic moment requires knowledge of the boosted wavefunction.  The
Melosh transformation is also essential for deriving the DHG sum
rule and low-energy theorems of composite systems \cite{bp69}.

A similar analysis can be performed for the axial-vector coupling
measured in neutron decay. The coupling $g_A$ is given by the
spin-conserving axial current $J_A^+$ matrix element
\begin{equation}
g_A(0) =\langle p,\uparrow | J^+_A | p,\uparrow \rangle\; .
\end{equation}
The value for $g_A$ can be written as $g_A=\langle \gamma_A \rangle
g_A^{\rm NR}$, with $g_A^{\rm NR}$ being the nonrelativistic value of
$g_A$ and with $\gamma_A$ given by \cite{cc91,ma91}
\begin{equation}
\gamma_A(x_i,k_{\perp i},m)=\frac{(m+x_3 {\cal M})^2-
k_{\perp 3}^2}{(m+x_3 {\cal M})^2+ k_{\perp 3}^2}\; .
\label{gammaa}
\end{equation}
In Fig.~\ref{fig3}(b) the axial-vector coupling is plotted against
the proton radius $M_p R_1$.  The same parameters and the same line
representation as in Fig.~\ref{fig3}(a) are used.  The functional
dependence of $g_A(M_p R_1)$ is also found to be independent of the
assumed wavefunction. At the physical proton radius $M_p R_1=3.63$,
one predicts the value $g_A = 1.25$ (indicated by the dotted lines
in Fig.~\ref{fig3}(b)), since $\langle \gamma_A \rangle =0.75$.  The
measured value is $g_A= 1.2573\pm 0.0028$ \cite{pdg92}.  This is a
25\%\ reduction compared to the nonrelativistic SU(6) value
$g_A=5/3,$ which is only valid for a proton with large radius $R_1
\gg 1/M_p.$ The Melosh rotation generated by the internal transverse
momentum \cite{ma91} spoils the usual identification of the
$\gamma^+ \gamma_5$ quark current matrix element with the total
rest-frame spin projection $s_z$, thus resulting in a reduction of
$g_A$.

Thus, given the empirical values for the proton's anomalous moment
$a_p$ and radius $M_p R_1,$ its axial-vector coupling is
automatically fixed at the value $g_A=1.25.$ This is an essentially
model-independent prediction of the three-quark structure of the
proton in QCD.  The Melosh rotation of the light-cone wavefunction
is crucial for reducing the value of the axial coupling from its
nonrelativistic value 5/3 to its empirical value. The near equality
of the ratios $g_A/g_A(R_1 \rightarrow \infty)$ and $a_p/a_p(R_1
\rightarrow \infty)$ as a function of the proton radius $R_1$ shows
the wave-function independence of these quantities.  We emphasize
that at small proton radius the light-cone model predicts not only a
vanishing anomalous moment but also $ \lim_{R_1 \rightarrow 0}
g_A(M_p R_1)=0$.  One can understand this physically: in the zero
radius limit the internal transverse momenta become infinite and the
quark helicities become completely disoriented.  This is in
contradiction with chiral models, which suggest that for a zero
radius composite baryon one should obtain the chiral symmetry result
$g_A=1$.

The helicity measures $\Delta u$ and $\Delta d$ of the nucleon each
experience the same reduction as does $g_A$ due to the Melosh
effect. Indeed, the quantity $\Delta q$ is defined by the axial
current matrix element
\begin{equation}
\Delta q=\langle p,\uparrow | \bar
q\gamma^+\gamma_5 q | p,\uparrow \rangle\; ,
\end{equation}
and the value for $\Delta q$ can be written analytically as $\Delta
q=\langle \gamma_A \rangle \Delta q^{\rm NR}$, with $\Delta q^{\rm
NR}$ being the nonrelativistic or naive value of $\Delta q$ and
$\gamma_A$ given by Eq. (\ref{gammaa}).

The light-cone model also predicts that the quark helicity sum
$\Delta\Sigma=\Delta u+\Delta d$ vanishes as a function of the
proton radius $R_1$. Since $\Delta\Sigma$ depends on the proton
size, it cannot be identified as the vector sum of the rest-frame
constituent spins. The rest-frame spin sum is not a Lorentz
invariant for a composite system \cite{ma91}. Empirically, one can
measure $\Delta q$ from the first moment of the leading-twist
polarized structure function $g_1(x,Q).$ In the light-cone and
parton model descriptions, $\Delta q=\int_0^1 dx [q^\uparrow (x) -
q^\downarrow (x)]$, where $q^\uparrow (x)$ and $q^\downarrow (x)$
can be interpreted as the probability for finding a quark or
antiquark with longitudinal momentum fraction $x$ and polarization
parallel or anti-parallel to the proton helicity in the proton's
infinite momentum frame  \cite{lb80}.  [In the infinite momentum
frame there is no distinction between the quark helicity and its
spin projection $s_z.$] Thus $\Delta q$ refers to the difference of
helicities at fixed light-cone time or at infinite momentum; it
cannot be identified with $q(s_z=+\frac{1}{2})-q(s_z=-\frac{1}{2}),$ the
spin carried by each quark flavor in the proton rest frame in the
equal-time formalism.

Thus the usual SU(6) values $\Delta u^{\rm NR}=4/3$ and $\Delta
d^{\rm NR}=-1/3$ are only valid predictions for the proton at large
$M R_1.$ At the physical radius the quark helicities are reduced by
the same ratio 0.75 as is $g_A/g_A^{\rm NR}$ due to the Melosh
rotation. Qualitative arguments for such a reduction have been given
elsewhere \cite{karl92,fritzsch90}.  For $M_p R_1 = 3.63,$ the
three-quark model predicts $\Delta u=1,$ $\Delta d=-1/4,$ and
$\Delta\Sigma=\Delta u+\Delta d = 0.75$.  Although the gluon
contribution $\Delta G=0$ in our model, the general sum rule
\cite{jm90}
\begin{equation}
\frac{1}{2}\Delta\Sigma +\Delta G+L_z= \frac{1}{2}
\end{equation}
is still satisfied, since the Melosh transformation effectively
contributes to $L_z$.

Suppose one adds polarized gluons to the three-quark light-cone
model. Then the flavor-singlet quark-loop radiative corrections to
the gluon propagator will give an anomalous contribution $\delta
(\Delta q)=-\frac{\alpha_s}{2\pi}\Delta G$ to each light quark
helicity \cite{Altarelli}.  The predicted value of $g_A = \Delta
u-\Delta d$ is of course unchanged. For illustration we shall choose
$\frac{\alpha_s}{2\pi}\Delta G=0.15$. The gluon-enhanced quark model
then gives the values in Table~1, which agree well with the present
experimental values. Note that the gluon anomaly contribution to
$\Delta s$ has probably been overestimated here due to the large
strange quark mass. One could also envision other sources for this
shift of $\Delta q$ such as intrinsic flavor \cite{fritzsch90}.  A
specific model for the gluon helicity distribution in the nucleon
bound state is given elsewhere \cite{bbs94}.

\begin{table}
\begin{center}
\begin{tabular}{|c|c|c|c|c|}
\hline Quantity & NR & $3q$ & $3q+g$ & Experiment \\ \hline $\Delta u$
& $\frac{4}{3}$ & 1 & 0.85 & $0.83\pm 0.03 $ \\ $\Delta d$
&$-\frac{1}{3}$
& $-\frac{1}{4}$ & --0.40 & $-0.43\pm 0.03 $\\ $\Delta s$ & 0 & 0 &
--0.15 & $-0.10\pm 0.03 $\\ $\Delta \Sigma$ &1 & $\frac{3}{4}$ & 0.30 &
$0.31\pm 0.07 $\\ \hline
\end{tabular}
\end{center}
\caption[*]{Comparison of the quark content of the proton
in the nonrelativistic quark model (NR), in the three-quark model
($3q$), in a gluon-enhanced three-quark model ($3q+g$), and with
experiment \cite{ek94}.
}
\end{table}

In the above analysis of the singlet moments, it is assumed that all
contributions to the sea quark moments derive from the gluon anomaly
contribution $\delta (\Delta q)=-\frac{\alpha_s}{2\pi}\Delta G$. In
this case the strange and anti-strange quark distributions will be
identical. On the other hand, if the strange quarks derive from the
intrinsic structure of the proton, then one would not expect this
symmetry.  For example, in the intrinsic strangeness wavefunctions,
the dominant fluctuations in the nucleon wavefunction are most
likely dual to intermediate $\Lambda$-$K$ configurations since they
have the lowest off-shell light-cone energy and invariant mass. In
this case $s(x)$ and $\bar s(x)$ will be different.

The light-cone formalism also has interesting consequences for spin
correlations in jet fragmentation.  In LEP or SLC one produces $s$
and $\bar s$ quarks with opposite helicity.  This produces a
correlation of the spins of the $\Lambda$ and $\overline\Lambda$,
each produced with large $z$ in the fragmentation of their
respective jet. The $\Lambda$ spin essentially follows the spin of
the strange quark since the $ud$ has $J=0$.  However, this cannot be
a 100\% correlation since the $\Lambda$ generally is produced with
some transverse momentum relative to the $s$ jet.  In fact, from the
light-cone analysis of the proton spin, we would expect no more than
a 75\% correlation since the $\Lambda$ and proton radius should be
almost the same. On the other hand if $z=E_\Lambda/E_s \to 1,$ there
can be no wasted energy in transverse momentum.  At this point one
could have 100\% polarization. In fact, the nonvalence Fock states
will be suppressed at the extreme kinematics, so there is even more
reason to expect complete helicity correlation in the endpoint
region.

We can also apply a similar idea to the study of the fragmentation
of strange quarks to $\Lambda$s produced in deep inelastic lepton
scattering on a proton.  One can  use the correlation
between the spin of the target proton and the spin of the $\Lambda$
to directly measure the strange polarization $\Delta s.$ It is
conceivable that any differences between $\Delta s$ and $\Delta \bar
s$ in the nucleon wavefunction could be distinguished by measuring
the correlations between the target polarization and the $\Lambda$
and $\overline\Lambda$ polarization in deep inelastic lepton proton
collisions or in the target polarization region in
hadron-proton collisions.

In summary, we have shown that relativistic effects are crucial for
understanding the spin structure of nucleons. By plotting
dimensionless observables against dimensionless observables, we
obtain relations that are independent of the momentum-space form of
the three-quark light-cone wavefunctions. For example, the value of
$g_A \simeq 1.25$ is correctly predicted from the empirical value of
the proton's anomalous moment. For the physical proton radius $M_p
R_1= 3.63$, the inclusion of the Wigner-Melosh rotation due to the
finite relative transverse momenta of the three quarks results in a
$\sim 25\%$ reduction of the nonrelativistic predictions for the
anomalous magnetic moment, the axial vector coupling, and the quark
helicity content of the proton.  At zero radius, the quark
helicities become completely disoriented because of the large
internal momenta, resulting in the vanishing of $g_A$ and the total
quark helicity $\Delta \Sigma.$

\section {CONCLUSIONS}

One of the central problems in particle physics is to determine the
structure of hadrons in terms of their fundamental QCD quark and
gluon degrees of freedom. As I have outlined in this talk, the
light-cone Fock representation of quantum chromodynamics provides
both a tool and a language for representing hadrons as fluctuating
composites of fundamental quark and gluon degrees of freedom. 
Light-cone quantization provides an attractive method to compute
this structure from first principles in QCD. However, much more
progress in theory and in experiment will be needed to fulfill this
promise.

\vspace{.5in}
\centerline{\Large\bf Acknowledgements}
Much of this talk is based on an earlier review written in
collaboration with Dave Robertson and on collaborations with 
Matthias Burkardt, Sid Drell, John Hiller, Kent Hornbostel, Paul
Hoyer, Hung Jung Lu, Bo-Qiang Ma. Al Mueller, Chris Pauli, Steve
Pinsky, Felix Schlumpf, Ivan Schmidt, Wai-Keung Tang, and Ramona
Vogt. I also thank Simon Dalley and Francesco Antonuccio for helpful
discussions.


\begin{thebibliography}{99}
\bibitem{r13} 
A summary of light-cone perturbative-theory calculation rules for
gauge theories is given by Lepage and Brodsky \cite{lb80}.  We
follow the notation of this reference here.
\bibitem{r15}
The polarization of each vector-boson constituent is specified by
the helicity index $S_i$ in $\psi$ as in Eq. (\ref{eq16}).
\bibitem{r17}
We use momentum conservation to eliminate the dependence of $\psi$
on $\longvec k\!\!_{\perp j}$, where $j$ is the struck quark.  Note
that the results (\ref{eq9}), (\ref{eq10}), and (\ref{eq14}) are
independent of the charge of the constituents.
\bibitem{r21}
For example, the contribution to the nucleon anomalous
moment in the quark model gives a contribution $\delta a \propto
gM_N/\left\langle k^2_\perp\right\rangle$ if there is a $g\phi^3$
trilinear coupling of scalars.  It is thus possible to obtain a
contribution to the anomalous moment of a fermion which is linear in
its mass even if all of its consituent fermions are massless.
\end{thebibliography}
\end{document}